\documentclass[12pt]{article}
\usepackage{amsmath}
\usepackage[dvips]{graphicx}
\usepackage{epsfig}
\usepackage{amsmath}
\usepackage{amssymb}
\usepackage{graphics,epstopdf}
\usepackage{amsfonts}
\usepackage{amsthm}
\usepackage[usenames]{color}

\definecolor{AV}{rgb}{0.65,0.0,0}
\definecolor{GC}{rgb}{0,0.0,0.65}
\definecolor{WS}{rgb}{0,0.65,0}

\usepackage[
      colorlinks=true,
      linkcolor=blue,
      urlcolor=blue,
      filecolor=blue,
      citecolor=red,
      pdfstartview=FitV,
      pdftitle={},
      pdfauthor={},
      pdfsubject={},
      pdfkeywords={},
      pdfpagemode=None,
      bookmarksopen=true
]{hyperref}

\newcommand{\bm}{\begin{multiline}}
\newcommand{\beq}{\begin{equation}}
\newcommand{\eeq}{\end{equation}}
\newcommand{\beqs}{\begin{eqnarray}}
\newcommand{\eeqs}{\end{eqnarray}}

\newcommand{\ra}{\rightarrow}

\begin{document}

\title{\textbf{\Large The Kiselev solution in power-Maxwell electrodynamics}}

\author{\textbf{Marina--Aura Dariescu}$^*$, \textbf{Ciprian Dariescu}$^*$, \textbf{Vitalie Lungu}$^*$\\  \textbf{and Cristian Stelea}$^{**}$ \\
$^{*}$Faculty of Physics, 
{ ``Alexandru Ioan Cuza"} University  of Iasi \\
Bd. Carol I, No. 11, 700506 Iasi, Romania \\
$^{**}$Department of Exact and Natural Sciences, \\
Institute of Interdisciplinary Research,\\
 { ``Alexandru Ioan Cuza"} University of Iasi, \\
 Bd. Carol I, no. 11, 700506 Iasi, Romania 
}

\date{}
\maketitle

\begin{abstract}
In this work we reconsider the solution describing black holes surrounded by a `quintessence'-like fluid. This geometry was introduced by Kiselev in 2003 and its physical source was originally modeled by an anisotropic fluid. We show that the Kiselev geometry is actually an exact solution of the Einstein equations coupled to nonlinear electrodynamics. More specifically, we show that the Kiselev geometry becomes an exact solution in the context of power-Maxwell electrodynamics, using either an electric ansatz or a magnetic one. In both cases the physical source can be modeled by a power-Maxwell Lagrangian, albeit with different powers corresponding to the electric or the magnetic charges. We briefly investigate the motion of charged particles in this geometry. Finally, we give the proper interpretation of the black-hole thermodynamics in this context. Similarly to the Schwarzschild-de Sitter case, we note the presence of the Schottky peaks in the heat capacity, signaling out the possibility of this thermodynamic black hole system to function as a continuous heat machine.
\end{abstract}

\begin{flushleft}
{\it Keywords}: Nonlinear electrodynamics; Power-Maxwell lagrangian; Quintessence fields
 \\
{\it PACS:}
04.20.Jb Exact solutions;
02.40.Ky Riemannian geometries;
04.62.+v Quantum fields in curved spacetimes;
02.30. Gp Special functions.
\end{flushleft}

\baselineskip 1.5em

\newpage

\section{Introduction}

Astrophysical observations from supernovae (Type Ia) \cite{Riess,Perlmutter}, cosmic microwave background radiation (CMBR) \cite{WMAP1,Planck} , Baryon acoustic oscillations (BAO) \cite{BAO1} and the Hubble measurements are suggesting an accelerating expansion of our universe, which may be explained by the presence of dark energy. The defining property of dark energy can be characterized by means of an effective equation of state parameterization of the form $P=w\rho$, where $P$ is the isotropic pressure, while $\rho$ is the energy density. For dark energy fluids one has to restrict $w<-\frac{1}{3}$. For example, a positive cosmological constant $\Lambda$ corresponds to a dark energy model for which $w=-1$ (see the review \cite{Padmanabhan:2002ji} and references therein).

One of the candidates for dark energy is the quintessence \cite{Capozziello:2005ra,Vikman:2004dc} (see also \cite{Tsujikawa:2013fta} and the references within), which can be seen as a canonical scalar field coupled to gravity whose potential is decreasing as the field increases. A slowly varying scalar field $\phi$ with an appropriate scalar potential $V(\phi)$ can lead to the accelerated expansion of the Universe \cite{Capozziello:2005mj}. At the cosmological level, the role of the scalar field has also been investigated in \cite{Dariescu:2004ex}, where the Einstein–Klein-Gordon equations for Friedmann–Robertson–Walker geometries were analyzed. However, it has been proved that the origin of quintessence at the cosmological and galaxy scales could be significantly different, in the sense that the quintessence state parameter $w$ can have different values \cite{Guzman:2000zba}.

Almost twenty years ago, a spherically-symmetric static solution of Einstein equations, describing black holes surrounded by `quintessence'-like fluids has been found by Kiselev \cite{Kiselev:2002dx}. The Kiselev geometry is sourced by an anisotropic fluid \cite{Visser:2019brz}, \cite{Boonserm:2019phw} that behaves like a kind of dark energy since its equation of state is $p_r=-\rho$, with $p_r\neq p_{t}$, where $p_r$ is the radial pressure, while $p_t$ denotes the tangential pressures $p_{\theta}$ and $p_{\varphi}$. The isotropic pressure is $P=\frac{1}{3}(p_r+2p_t)=w\rho$, where $w$ is the Kiselev quintessence parameter. Therefore, in order to cause the accelerated expansion of the universe, the equation of state parameter $w$ should be in the range $w \in [ -1 , -1/3]$. In the particular case $w = -2/3$, one gets in the metric function an additional linear contribution, $-kr$. The parameter $k$, which is the Kiselev quintessence charge can lead to a metric that was previously used in modified Newtonian dynamics (MOND) \cite{Mannheim}, and various other modifications of General Relativity \cite{Gregoris:2021plc},\cite{Panpanich:2018cxo} (see also \cite{Toledo} and \cite{Dariescu:2022vhw}).

In this paper we propose a physical source for the Kiselev geometry in the context of nonlinear electrodynamics. This class of theories has a long history, as they were introduced initially in order to cure the infinite electric field and the infinite self-energy for point-like charged particles. The first model was introduced by Born and Infeld in 1934 \cite{Born} and it was soon realized that nonlinear electrodynamics theories could act as effective classical modifications from QED \cite{Euler}. Nowadays, the nonlinear electrodynamics theories are an active area of research and they provided us with various interesting black hole solutions in four and higher dimensions (for a review and more references see \cite{Breton:2007bza} and \cite{Bokulic:2021dtz}). Generically, the nonlinear Lagrangian is constructed from the two quadratic electromagnetic invariants $F_{\mu\nu}F^{\mu\nu}$ and $F_{\mu\nu}\star F^{\mu\nu}$, where $\star F^{\mu\nu}$ is the dual of $F_{\mu\nu}$ \cite{Sorokin:2021tge} (see also \cite{Gibbons:2001sx}). Here we will focus on a simpler Lagrangian $L=L(F)$ that depends only the first electromagnetic invariant, $F=F_{\mu\nu}F^{\mu\nu}$. From the least action principle we then obtain the following field equation for the nonlinear electromagnetic field:
\beqs
\partial_{\mu}\left(\sqrt{-g}L_FF^{\mu\nu}\right)&=&0,
\eeqs 
where $L_F$ is the derivative of the function $L$ with respect to the variable $F$.

Nonlinear electrodynamics has been used previously as a source for the so-called Bardeen regular black holes (see for instance \cite{Flachi:2012nv} and \cite{Ayon-Beato:2000mjt}).

More recently, an exact solution involving a regular Bardeen black hole with quintessence was previously found in \cite{Rodrigues:2022qdp}. The results of that paper could hint towards a possible explanation of the quintessence fields in terms of non-linear electromagnetic fields. However, the limit $q\ra 0$ in which this solution reduces to the Kiselev black hole is problematic since it would not be a solution of the nonlinear electromagnetic field anymore as the nonlinear electromagnetic field is sourced by the magnetic monopole charge $q$. Moreover, the Lagrangian found in \cite{Rodrigues:2022qdp} looks more like an on-shell Lagrangian for the nonlinear electromagnetic field of the Bardeen-Kiselev solution, since it depends explicitly on the parameters of that particular solution (such as the mass $M$, charge $q$ and the quintessence parameters $c$\footnote{The parameter $c$ in \cite{Rodrigues:2022qdp} corresponds to the parameter $k$ in our paper.} and $\omega$). This approach will offer no clues regarding the electromagnetic origins of the Kiselev parameters $c$ and $\omega$ as they cannot be re-interpreted in terms of quantities related to the nonlinear electromagnetic fields.

In our work we will consider a different approach, in context of the so-called power-Maxwell model, for which the electromagnetic Lagrangian is defined as $-\alpha (F_{\mu\nu}F^{\mu\nu})^q$, where $\alpha$ is a coupling constant that has to be introduced in order to have a positive energy density of the nonlinear Maxwell field \cite{Hassaine:2008pw}, \cite{Gonzalez:2009nn}. In spaces with $d$ dimensions it turns out that the power-Maxwell electrodynamics can still enjoy conformal invariance if the power coefficient $q$ in the Lagrangian is equal to $\frac{d}{4}$ \cite{Hassaine:2007py}. For more properties of the solutions of the power-Maxwell theory in various theories and various dimensions see also \cite{EslamPanah:2021xaf} - \cite{Hendi:2010zza}.

Within this nonlinear electrodynamics theory, we show that the Kiselev solution becomes an exact solution of the Einstein-power-Maxwell equations (with or without cosmological constant) using either an ansatz involving electric charges and fields, or a magnetic monopole ansatz. In both cases we show explicitly how one can relate the Kiselev quintessence parameters $k$ and $\omega$ to the corresponding electric charge $Q_e$ or the magnetic charge $Q_m$ and the power coefficient $q$ appearing in the power-Maxwell Lagrangian. We found the interesting fact that the power $q$ corresponding to an electric charge and the power $q$ for a magnetic charge differ considerably (while they agree for $q=1$, that is for the usual Maxwell electrodynamics). This might signal the absence of dyonic black holes in these theories and probably there will be problems when dealing with the rotating versions of these geometries.

We then describe some of the properties of these solutions: we study the motion of charged and uncharged particles around these black holes, noting that light will travel on the null-geodesics defined by an ``effective geometry'', instead of the null geodesics of the background geometry \cite{Novello:1999pg}, \cite{Breton:2007bza}, \cite{Habibina:2020msd}.

The motion of particles moving around different types of black holes surrounded by quintessence has been extensively studied in recent years. In this regard, for the Schwarzschild black hole surrounded by quintessence, the null and timelike geodesics
have been investigated by many authors, see for instance \cite{Fernando:2012ue}-\cite{Al-Badawi}. In the present paper, we are discussing the effective potential for charged particles moving around the black hole described by Kiselev's solution as an exact solution of the power-Maxwell nonlinear electrodynamics. By comparing the results with those for the usual Schwarzschild-Kiselev black hole, one can point out the effects of the nonlinear electromagnetic fields.

The structure of this paper is as follows: in the next section we review the properties of the Kiselev solution that describes a black hole surrounded by a quintessence-like anisotropic fluid. In section \ref{section3} we introduce the power-Maxwell theory and show that the Kiselev solution becomes an exact solution in this theory. In section \ref{section4} we discuss the geodesic motion for timelike and null cases We also present the effective potential for charged particles moving around the black hole in this geometry. In section \ref{section5} we approach the thermodynamic properties of the Kiselev black holes. In particular, we point out the presence of the Schottky peaks in the heat capacity \cite{Dinsmore:2019elr}, \cite{Johnson:2019ayc}, \cite{Johnson:2019vqf} which hints to the possibility of interpreting the Kiselev black hole as a continuous heat machine. Section \ref{section6} is dedicated to conclusions and avenues for further work.

\section{The Kiselev geometry}
\label{section2}

The Kiselev geometry is described by the following static four-dimensional line-element  \cite{Kiselev:2002dx}:
\begin{equation}
ds^2= - g(r)dt^2+\frac{dr^2}{g(r)} + r^2 ( d \theta^2 + \sin^2 \theta d \varphi^2)
\label{kiselev} \; ,
\end{equation}
where\footnote{In some works the parameter $k$ is denoted as $c$.}
\begin{equation}
g(r) = 1 -\frac{2M}{r} - \frac{k}{r^{3 w +1}}.
\label{g0}
\end{equation}
Here $w$ is the equation of state parameter and $k$ is a positive quintessence parameter, which is related to the fluid quintessence energy density $\rho$:
\[
\rho = - \frac{3kw}{r^{3(w+1)}} \; ,
\]
while components of the anisotropic pressures can be written as $p_r=-\rho$ and the tangential pressures are given by:
\beqs
p_{\theta}&=&p_{\varphi}=-\frac{3(3w+1)kw}{2r^{3(w+1)}}.
\eeqs
In order to have an accelerated expansion, the equation of state parameter $w$ should belong to the interval $w \in [ -1 , -1/3]$.

\begin{figure}
  \centering
  \includegraphics[width=0.45\textwidth]{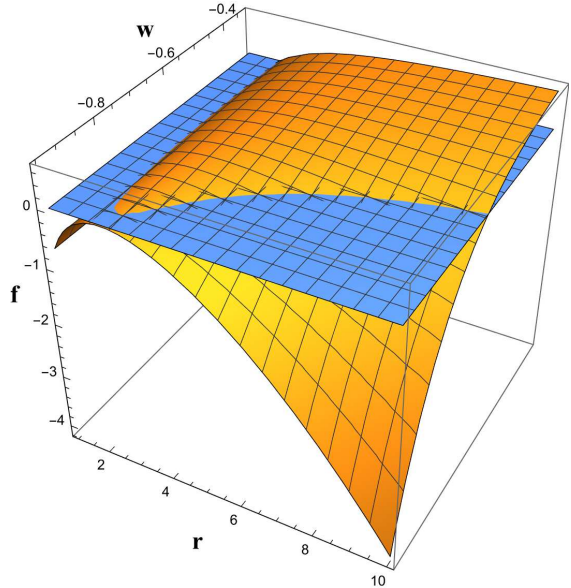}
  \caption{The metric function $f (r) = g(r)$ given in (\ref{g0}), for $w \in [-1/3 , -1]$ with $M=0.8$ and $k=0.2$. The blue plane corresponds to $g(r) =0$.} 
  \label{figure1}
\end{figure}

The horizons of this geometry will correspond to solutions of the equation $g(r)=0$. As it can be noticed in Figure $\ref{figure1}$, for fixed values of $M$ and $k$, one may have, besides the black hole's horizon $r_b$, situated just after $2M$, an additional (outer) quintessence horizon, $r_q$. The two horizons are given by the intersection between the $f(r) = g(r)$ sheet and the $r=0$ plane. Once $w$ is decreasing to the limiting value $w \to -1$, the quintessence horizon comes closer to the black hole's horizon.
In the opposite case where $w$ is increasing to $w \to -1/3$, the quintessence horizon moves to bigger values of the radial coordinate.
For $w=-1/3$, we have a static black hole surrounded by a spherically symmetric cloud of strings, with 
\begin{equation}
g(r) = 1-k - \frac{2M}{r}  \; ,
\end{equation}
and the unique horizon
\begin{equation}
r_b = \frac{2M}{1-k} \; .
\label{rbp0}
\end{equation}
This spacetime is still singular as it contains conical singularities. There is no quintessence horizon, the effect of the quintessence fluid being encoded only in the constant $k$, which is responsible for the deficit or the excess of solid angle.
 
For the limiting value $w =-1$, one recovers the Schwarzschild–de Sitter geometry, with
\[
g(r) = 1-\frac{2M}{r} - k r^2\, .
\]
In this case, three real roots of the corresponding cubic equation, among which two are positive, are obtained for the relation between the parameters $27 k M^2  <1$. Otherwise, one has one real root and two complex conjugated roots.

The physically important case corresponding to $w =-2/3$ is particularly simple, since one has a linear contribution in the metric function, which becomes:
\begin{equation}
g(r) = 1 -\frac{2M}{r}  - k r \, .
\label{w23}
\end{equation}
The two horizons, solution of the equation $g(r) = 0$, are obtained for $8kM < 1$ and they are given by:
\begin{equation}
r_{\pm} = \frac{1}{2k} \left[ 1 \pm \sqrt{1-8kM} \right]  .
\label{rpm}
\end{equation}

However, note that if $M=0$, \textit{i.e.} in absence of the black hole horizon, while the geometry still has a quintessence horizon, located at $r_+$, there is a naked curvature singularity in origin unless $w=-1$ and $w=-\frac{1}{3}$. In this case the quintessence horizon plays the role of a cosmological horizon, similar to the de Sitter case.

\section{The Kiselev solution in power-Maxwell electrodynamics}
\label{section3}

One interesting class of nonlinear electromagnetic sources is the power-Maxwell theory. The full action in this case is given by \cite{Hassaine:2008pw}, \cite{Gonzalez:2009nn}:
\beqs
I=-\frac{1}{16\pi G}\int_{\cal V}d^4x\sqrt{-g}\left(R-\alpha F^q\right)-\frac{1}{8\pi G}\int_{\partial {\cal V}}d^3x\sqrt{-\gamma}K+I_{bd},
\label{actionfull}
\eeqs
where we denoted $F=F_{\mu\nu}F^{\mu\nu}$ and K is the usual Gibbons-Hawking boundary term, defined on the spacetime boundary $\partial {\cal V}$, on which the induced metric is denoted by $\gamma_{ab}$. The terms $I_{bd}$ refer to possible counterterm-like terms (for the gravitational and/or electromagnetic fields) needed to render the full action (\ref{actionfull}) finite.

The field equations derived from this action can be written in the form:
\beqs
G_{\mu\nu}=T_{\mu\nu}\\
\partial_{\mu}\left(\sqrt{-g}F^{\mu\nu}F^{q-1}\right)&=&0,
\label{eom}
\eeqs
where the stress-energy tensor of the electromagnetic field is defined as:
\beqs
T_{\mu\nu}&=&2\alpha\bigg[qF_{\mu\rho}F_{\nu}^{~\rho}F^{q-1}-\frac{1}{4}g_{\mu\nu}F^q\bigg].
\label{tem}
\eeqs
We are looking for a spherically-symmetric geometry of the Kiselev form:
\beqs
ds^2&=&-f(r)dt^2+\frac{dr^2}{f(r)}+r^2(d\theta^2+\sin^2\theta d\varphi^2)
\label{metric}
\eeqs
where
\beqs
f(r)&=&1-\frac{2M}{r}-kr^p.
\label{fr}
\eeqs
Here $k$ is a positive parameter, which can be related to the original quintessence parameter in Kiselev's solution, while $p=-(3w+1)$ is now a positive parameter. If $w\in[-1 , -\frac{1}{3}]$ then $0\leq p\leq 2$, although in the general solution one can keep $p$ general.

\subsection{The electric ansatz}

If the geometry is sourced by nonlinear electric fields we shall use the following ansatz for the electromagnetic potential:
\beqs
A_{\mu}&=&\left(\chi(r), 0, 0, 0\right),
\label{ela}
\eeqs
from which one constructs the Maxwell tensor $F_{\mu\nu}=\partial_{\mu}A_{\nu}-\partial_{\nu}A_{\mu}$. A quick computation reveals that the electromagnetic invariant takes the form $F=-2\left(\frac{d\chi}{dr}\right)^2<0$, which allows us to pick $\alpha=(-1)^{(-q)}$ in the action, in order to have a positive-definite energy of the electromagnetic field for $q<0$. More specifically, if one computes the components of the electromagnetic stress-energy tensor (\ref{tem}), then the energy density of the electromagnetic field is:
\beqs
\rho&=&-T^t_t=\frac{\alpha F^q (1-2q)}{2}.
\eeqs
One can see that one can choose $\alpha=(-1)^{-q}$ if $q<\frac{1}{2}$ and $\alpha=-(-1)^{-q}$ if $q>\frac{1}{2}$ if the energy density $\rho$ is to be positive-definite. In our case, for the Kiselev solution with quintessence we will see bellow that the power parameter $q$ can take only negative values, such that $\alpha=(-1)^{-q}$ is the appropriate choice.

One can now solve directly the equation of motion for the nonlinear electromagnetic field in (\ref{eom}) and one obtains:
\beqs
\chi(r)&=&C_1+C_2r^{p+1},
\label{chiel}
\eeqs
where $C_1$ and $C_2$ are constants of integration. 

Solving now the Einstein equations in (\ref{eom}) one can further find that the value of the power coefficient $q$ must be restricted in terms of the parameter $p$ as\footnote{Note that for $0\leq p\leq 2$ the power $q<0$ such that $\alpha=(-1)^{-q}$ is the right choice.} :
\beqs
q&=&\frac{p-2}{2p},
\eeqs
while the quintessence parameter $k$ in (\ref{fr}) can be written as\footnote{More generally, $k$ contains the factor $\alpha(-1)^q$ multiplying the expression given in (\ref{kel}). However, this factor cancels out for $\alpha=(-1)^{-q}$. This factor is responsible for changing the sign of $k$ in the usual Maxwell electrodynamics, when $q=1>\frac{1}{2}$ and $\alpha=-(-1)^{-q}$ accordingly.}:
\beqs
k&=&\frac{2^{\frac{1}{2}-\frac{2}{p}}C_2^{1-\frac{2}{p}}(p+1)^{-\frac{2}{p}}}{p}.
\label{kel}
\eeqs

The constant $C_2$ can be directly related to the electric charge $Q_e$ of this black hole solution. The electric charge can be computed using the formula:
\beqs
Q_e&=&-\frac{\alpha}{4\pi}\int_{S^2}(F)^{q-1}\star F=2^{-\frac{p+2}{2p}}C_2^{-\frac{2}{p}}(p+1)^{-\frac{2}{p}},
\label{qel}
\eeqs
where $\star F_{\mu\nu}=\frac{1}{2}\sqrt{-g}\epsilon_{\mu\nu\alpha\beta}F^{\alpha\beta}$ is the dual of the Maxwell tensor $F_{\mu\nu}$, with $\epsilon_{\mu\nu\alpha\beta}$ being the Kronecker symbol, while $S^2$ is a $2$-sphere containing the black hole horizon.

Combining now (\ref{kel}) and (\ref{qel}) one can re-express the constant $C_2$ in (\ref{chiel}) as
\beqs
C_2&=&\frac{Q_e^{-\frac{p}{2}}}{2^{\frac{p+2}{4}}(p+1)}
\eeqs
Finally, the parameter $k$ appearing in the metric (\ref{fr}) can be expressed in terms of the electric charge $Q_e$ in the particularly simple form:
\beqs
k&=&\frac{2^{\frac{2-p}{4}}Q_e^{\frac{2-p}{2}}}{p(p+1)}.
\eeqs
Note that the obtained solution is valid for every value of the parameter $p$ and it can be easily modified to accommodate the presence of a cosmological constant $\Lambda$ in (\ref{actionfull}) and (\ref{fr}). 

\subsubsection{The energy conditions using the electric ansatz}

One can now easily check the energy conditions satisfied by the nonlinear electromagnetic field in our solution. Recall that in the original Kiselev solution with quintessence the parameter $p\in[0,2]$ such that $q\leq0$. More specifically, if one denotes the effective tangential pressures in the electromagnetic stress-energy tensor  (\ref{tem}) by:
\beqs
p_{\theta}=p_{\varphi}=-p_0,
\eeqs
then one can express the energy density $\rho$ and the radial pressure $p_r$ as:
\beqs
\rho=-p_r=p_0(1-2q),
\eeqs
where we defined
\beqs 
p_0=\frac{\big[2\left(\frac{d\chi}{dr}\right)^2\big]^q}{2}>0.
\eeqs
It is now easy to check that the Weak Energy Condition (WEC), which requires $\rho\geq 0$ and $\rho+p_r\geq 0$ and $\rho+p_t\geq0$ is satisfied for $q\leq 0$. Similarly, the Dominant Energy condition (DEC), which requires that $\rho\geq 0$ and $-\rho\leq p_r\leq \rho$ and $-\rho\leq p_t\leq \rho$  is satisfied as well. However, the Strong Energy Condition (SEC) is not satisfied since $\rho+p_r+2p_t<0$ in our case.

\subsection{The magnetic ansatz}

It is interesting to note that the same geometry defined by (\ref{fr}) can also be sourced by using a magnetic monopole ansatz for the electromagnetic potential. More specifically, we shall choose:
\beqs
A_{\mu}&=&(0, 0, 0,Q_m(1-\cos\theta)),
\eeqs
where $Q_m$ is a constant that can be shown to be equal to the magnetic charge in our solution. Indeed, in this case the magnetic monopole charge is defined by the equation:
\beqs
Q_m&=&\frac{1}{4\pi}\int_{S^2} F,
\label{qmag}
\eeqs
where $F=Q_m\sin\theta d\theta\wedge d\varphi$ denotes here the electromagnetic 2-form, not to be confused with the electromagnetic invariant $F=F_{\mu\nu}F^{\mu\nu}$ that we used in the action (\ref{actionfull}). In fact, it can be shown by direct computation that this invariant takes now the particularly simple form $F=\frac{2Q_m^2}{r^4}\geq 0$. Since this invariant is always positive, one can simply take $\alpha=1$ in the action (\ref{actionfull}) since we are guaranteed that the energy density of the electromagnetic field is now positive for any power coefficient $q$. Indeed, if one computes the corresponding energy density of the electromagnetic field for a magnetic charge one finds the explicit expression:
\beqs
\rho&=&\frac{\alpha F^q}{2}
\eeqs
and since $F>0$ one is forced to choose $\alpha=1$ for every value of $q$. One should contrast this to the results in the electric case, where the value of $\alpha$ depends on the value of $q$ and the sign of $F$. This might make problematic the construction of dyonic black holes in nonlinear power-Maxwell electrodynamics!

Moreover, with this magnetic ansatz the nonlinear Maxwell equations in (\ref{eom}) are satisfied as well.

Solving now the Einstein equations in (\ref{eom}) one finds that they are identically satisfied if one takes the following parameters:
\beqs
q&=&\frac{2-p}{4}
\label{qmagnetic}
\eeqs
in the action (\ref{actionfull}) and
\beqs
k&=&\frac{\left(2Q_m^2\right)^{\frac{2-p}{4}}}{2(p+1)}
\label{kmag}
\eeqs
in the metric (\ref{fr}). This completes our solution using the magnetic ansatz. Note that if $p\in[0,2]$, then the power coefficient $q$ belongs to the interval $0\leq q\leq \frac{1}{2}$.

\subsubsection{The energy conditions using the magnetic ansatz}

One can now discuss the energy conditions satisfied by our nonlinear electromagnetic field using the magnetic ansatz. If one considers the diagonal components of the corresponding stress-energy tensor (\ref{tem}):
\beqs
T^{\mu}_{~\nu}&=&diag (-\rho, p_r, p_{\theta}, p_{\varphi})
\eeqs
then one finds $\rho=-p_r=\frac{2p_0}{p}$ and $p_{\theta}=p_{\varphi}=-p_0$, where we defined $p_0$ by:
\beqs
p_0&=&\frac{p\left(\frac{2Q_m^2}{r^4}\right)^q}{4}.
\eeqs
Note that for the Kiselev with quintessence geometry one has $0\leq p\leq 2$ and therefore the quantity $p_0$ is positive.

One can now easily check that the weak energy condition (WEC) and the dominant energy condition (DEC) are both satisfied for $0\leq p\leq 2$, while the strong energy condition (SEC) is violated since $\rho+p_r+p_{\theta}+p_{\varphi}=-2p_0<0$, just as in the electric case.

\subsection{The location of the horizons}
\label{horizonsg}

We shall now describe some of the properties of the solutions found in the power-Maxwell nonlinear electrodynamics. First, let us discuss the location of the horizons in the geometry defined by (\ref{fr}). Recall that in our case $p\in[0,2]$ and $k>0$. 

\begin{figure}
  \centering
  \includegraphics[width=0.45\textwidth]{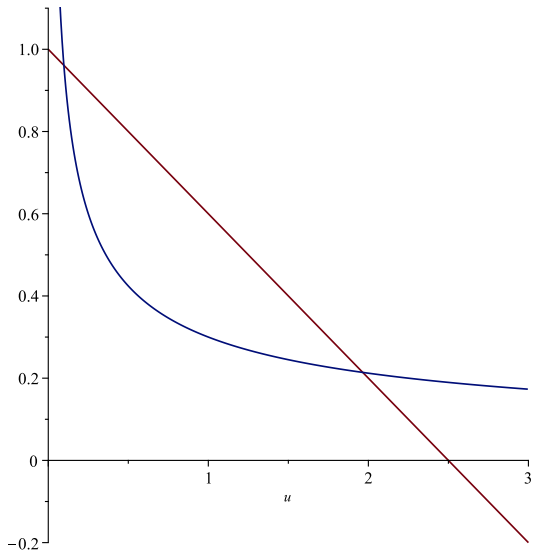}
  \caption{The intersection of the line described by $y_1$ for $M=0.2$ and the curve $y_2$ for $k=0.3$ and $p=0.5$.} 
  \label{figure2}
\end{figure}

The horizons will correspond to solutions of the equation $f(r)=0$. Even if, for general powers $p$, one cannot solve this equation analytically, it is still possible to draw some general conclusions regarding the location of the horizons. For this, let us define a new variable $u=\frac{1}{r}$ such that $f(u)=1-2Mu-ku^{-p}$. Then the solutions of the equation $f(u)=0$ will correspond to the intersection points of the line $y_1=1-2Mu$ with the curve $y_2=ku^{-p}$ (see Figure \ref{figure2}). One can see that depending on the values taken by $M$, $p$ and $Q$ one can have at most two horizons: a black hole horizon located at $r_b$ and one horizon (the quintessence horizon in the original Kiselev solution) located at $r_c>r_b$. This second horizon plays the role of an effective cosmological horizon and the static patch of the Kiselev geometry is now restricted to values of the radial coordinate $r$ in between these two horizons. Outside the `cosmological' horizon $r_c$ the radial coordinate becomes timelike and the temporal coordinate becomes spacelike, just as it happens in the Schwarzschild-de Sitter geometry. There is however a marked difference to the Schwarzschild-de Sitter case, since in absence of the black hole horizon there is a naked singularity in the bulk, unless $p=2$ (which corresponds to an effective de Sitter geometry) or $p=0$, which corresponds to a conical singularity.

As one increases the black hole mass parameter, $M$, the black hole horizon $r_b$ increases while the cosmological horizon $r_c$ shrinks. For fixed $Q$ and $p$ values there is a maximum value of the mass parameter $M$ such that the two curves $y_1$ and $y_2$ have only one point of intersection, located at 
\beqs
r_0&=&\frac{2(p+1)M}{p}.
\label{r0}
\eeqs
This situation corresponds to the extremal case, similar to the Nariai solution in the Schwarzschild-de Sitter case. The maximum value of the mass parameter can be expressed as:
\beqs
M_{max}&=&\frac{1}{2}p(p+1)^{\frac{p-1}{2-p}}k^{\frac{1}{2-p}},
\label{mmax}
\eeqs
where $k$ is given by (\ref{kel}) in the electric case and (\ref{kmag}) in the magnetic solution.

If the mass parameter is increased beyond this value the spacetime becomes singular, as it contains now a naked singularity. The lesson to be learnt is that the black hole horizon $r_b$ can only have values in between $0$ and $r_0$, otherwise the spacetime geometry is singular.

\section{Particle orbits in the power-Maxwell Kiselev background}
\label{section4}

In this section we shall briefly discuss the trajectories of particles in the Kiselev geometry as parameterized in (\ref{fr}). These trajectories have been previously discussed in literature in the timelike and the null cases \cite{Fernando:2012ue} -  \cite{Uniyal:2014paa}. One can express the parameters of the original Kiselev geometry in terms of our re-parameterized solution as $k>0$, while $p=-(3\omega+1)$. From this point of view we do not expect to find new results concerning the geodesics of the original Kiselev geometry. However, since our solution is an exact solution of a nonlinear electrodynamics theory, one should note that the test charged particles should couple to the nonlinear electromagnetic fields and therefore they will be influenced by these fields and their motion will change accordingly. In this section we shall describe the effects of the nonlinear electromagnetic fields for charged test particles using an effective potential method.

 However, as is now known in nonlinear electrodynamics theories, one should note that photons will follow the null geodesics of an ``effective'' geometry instead of the null geodesics of the original Kiselev background \cite{Novello:1999pg}. This will signal a new effect for photon's propagation due to our reinterpretation of the Kiselev geometry in terms of nonlinear electromagnetic fields.

\subsection{Timelike trajectories}

The trajectory of a massive test particle with mass $m$ and charge $e$ can be determined by using the Lorentz-force equation. In this case it reads:
\beqs
\frac{d^2x^{\mu}}{d\tau^2}+\Gamma^{\mu}_{\alpha\beta}\frac{dx^{\alpha}}{d\tau}\frac{dx^{\beta}}{d\tau}=\frac{e}{m}F^{\mu}_{~~\sigma}\frac{dx^{\sigma}}{d\tau},
\eeqs
where $\tau$ is the proper time along its trajectory. The proper time $\tau$ is defined by using the equation:
\begin{equation}
 -d \tau^2=- f(r) dt^2+\frac{dr^2}{f(r)} + r^2 (  d \theta^2 + \sin^2 \theta d \varphi^2 )  \; ,
\label{tau}
\end{equation}

The Lorentz equation can be derived from the Lagrangian corresponding to a charged particle moving in an electromagnetic field defined by $A_{\mu}$, given in (\ref{ela}) in the electric case:
\beqs
{\cal L}&=&\frac{1}{2}g_{\mu\nu}u^{\mu}u^{\nu}+\frac{e}{m}u^{\mu}A_{\mu}.
\eeqs

For a test uncharged particle ($e=0$) which is freely falling in the equatorial plane ($\theta = \pi /2$), the conserved energy $E$
and angular momentum $L$ are given by
\begin{equation}
E=f(r) \dot{t} \; , \quad L =  r^2 \dot { \varphi } \; ,
\label{ELn}
\end{equation}
where {\it dot} means the derivative with respect to $\tau$.
Using the two conserved quantities $E$ and $L$ along the trajectory, one can re-express the equation (\ref{tau}) as describing the motion of particle in an effective potential:
\begin{equation}
\dot{r}^2 +f(r) \left[ 1 + \frac{L^2}{r^2} \right] -E^2=0,
\label{e1}
\end{equation}
pointing out the effective potential
\begin{equation}
V_{eff} =f(r)\left[ 1 + \frac{L^2}{r^2} \right] = \left[ 1 -\frac{2M}{r} - kr^p \right] \left[ 1 + \frac{L^2}{r^2} \right],
\label{poteff}
\end{equation}
that is felt by an uncharged test particle. As expected, this effective potential is the same with the one previously derived in literature \cite{Fernando:2012ue} -  \cite{Uniyal:2014paa} for the original Kiselev geometry once one notices that $p=-(3\omega+1)$, while $k=c>0$ is the same parameter as in the original Kiselev geometry. 

The situation is more complicated for an electrically charged particle. In the original Kiselev geometry the electric charge of the test particle would have no effect on its trajectory, while in our reinterpreted Kiselev solution in terms of the nonlinear electromagnetic fields the situation changes. More specifically, in this case, while the expression for the angular momentum $L$ remains the same as in (\ref{ELn}), the expression of the energy $E$ is changed to take into account the effect of the electric charge of the particle moving in an nonlinear electromagnetic field:
\beqs
E&=&f(r)\dot{t}+\frac{e}{m}A_t,
\eeqs
where $A_t=\frac{Q_e^{-\frac{p}{2}}}{2^{\frac{p+2}{4}}(p+1)}r^{p+1}$ is the electric potential of the nonlinear electromagnetic field (\ref{ela}). One can now re-express the equation (\ref{tau}) in the form:
\beqs
\dot{r}^2&=&\left(E-\frac{e}{m}A_t\right)^2-f(r)\left(1+\frac{L^2}{r^2}\right)=(E-V_+)(E-V_-),
\eeqs
where we defined\footnote{For the corresponding analysis in the Reissner-Nordstrom case see \cite{Pugliese:2011py}.}:
\beqs
V_{\pm}&=&\frac{e}{m}A_t\pm\sqrt{f(r)\left(1+\frac{L^2}{r^2}\right)}
\label{vpcharge}
\eeqs
Note that it is the potential $V_+$ which corresponds to future-directed orbits for the charged particles. For an uncharged particle $e=0$ one recovers the effective potential (\ref{poteff}) as $V_{eff}=V_+V_-$.

Consider now again the case of an uncharged test particle with zero angular momentum $L=0$ (a test particle in radial fall) for which the relations (\ref{e1}) and (\ref{poteff}) turn into
\begin{equation}
\dot{r}^2 = E^2 - f(r) \, ,
\label{L0}
\end{equation}
and
\begin{equation}
V_{eff} = 1 -\frac{2M}{r} - kr^p
\end{equation}
and
the `effective' force acting on the particle has the expression
\[
F = - \frac{1}{2}V_{eff}^{\prime} = - \frac{1}{r^2} \left[ 2M -pkr^{p+1} \right] .
\]
The factor $\frac{1}{2}$ above appears once we take into account the fact that the equation of motion is derived from (\ref{L0}). Note that in our reinterpreted Kiselev geometry the parameter $k>0$ is now related to the electric charge $Q_e$ of the nonlinear electromagnetic field.

This effective force is attractive for $p\leq 0$, as in the Schwarzschild case, and it can have both attractive and repulsive contributions for $p>0$. The maxima of the effective potential are found by solving $F=0$ and one finds the root $r_0=\left(\frac{2M}{kp}\right)^{\frac{1}{p+1}}$. The maximum value of the effective potential is :
\beqs
V^{max}_{eff}&=&1-\frac{2M}{r_0}\frac{p+1}{p}.
\eeqs

\begin{figure}
  \centering
  \includegraphics[width=0.45\textwidth]{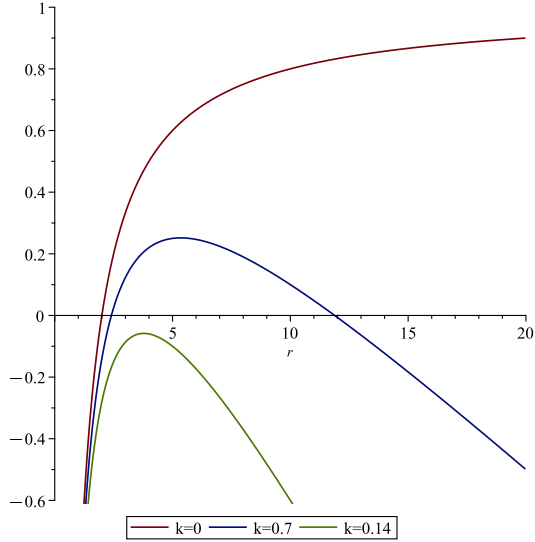}
  \caption{The effective potential given by (\ref{Vp1}) with $M=1$ for the Schwarzschild black hole, \textit{i.e.} $k=0$, and for the power-Maxwell black hole with $k=0.07$, respectively for $k=0.14$.} 
  \label{figure3}
\end{figure}

As an example, in Figure \ref{figure3} , we have represented the effective potential (\ref{poteff}) for $p=1$, i.e.
\begin{equation}
V_{eff} = 1 - \frac{2M}{r} - kr,
\label{Vp1}
\end{equation}
for different values of the parameter $k$. The usual Schwarzschild black hole, i.e. $k=0$, is represented by the red line, while the other solid lines correspond to the black hole in the nonlinear electrodynamics for nonzero values of $k$.
One may notice that, for $k< 1/(8M)$, the potential has a positive maximum, $V_{max} = 1-\sqrt{8kM}$, in $r_0= \sqrt{2M/k}$, and therefore there are two solutions of the equation $V_{eff}=0$, given by the intersection of the solid curve with the horizontal axis.
Once $k$ is increasing, the curve is approaching the horizontal axis. For $8kM>1$, the maximum of the potential moves below the horizontal axis, pointing out the formation of a naked singularity.

In order to have a stationary ``orbit'' (since $L=0$ one considers a stationary particle at a fixed radial position), with $r=R_c$, one has to impose the conditions
$\dot{r}=0$ and $\ddot{r}=0$, i.e.
\begin{equation}
V_{eff} (r= R_c ) = E^2 \; , \; \; V_{eff}^{\prime} ( r= R_c ) =0 \, .
\label{circular}
\end{equation}
For an arbitrary parameter $p$, the second equation in (\ref{circular}) leads to the radius
\begin{equation}
R_c =r_0=\left(\frac{2M}{kp}\right)^{\frac{1}{p+1}}
\end{equation}
and one may notice that for $p<0$ there are no stationary orbits.
If $p=2$, the stationary orbit has the radius
\[
R_c = \left( \frac{M}{k} \right)^{1/3}
\]
and the corresponding effective potential at $r=R_c$ reads
\[
V_{eff}= 1 - 3 (kM^2)^{1/3} \; .
\]
Since $V_{eff}^{\prime \prime} = - 6k <0$,
this stationary orbit is unstable.

In the physically important case $p=1$, the stationary radius is
\begin{equation}
R_c = \sqrt{\frac{2M}{k}} \; , 
\end{equation} 
and the effective potential reaches a maximum value
\begin{equation}
V_{eff} (r=R_c) = 1 - \sqrt{8kM} = V_{max} \; .
\end{equation}
Thus, the particle with $E^2 =  1 - \sqrt{8kM}$
has an unstable stationary orbit \cite{Fernando:2012ue}, \cite{Al-Badawi}.

As expected, the situation is more complicated when one deals with charged test particles in nonlinear electromagnetic fields. In this case, the circular motion is determined by the effective potential $V_+$ given in (\ref{vpcharge}). Circular orbits are determined again by the conditions $\dot{r}=0$ and $\ddot{r}=0$, which lead in our case to:
\beqs
E&=&V_+,~~~~~~~~\frac{dV_+}{dr}=0.
\label{circ}
\eeqs
One could solve, for instance the second equation to find the angular momentum $L$ of the charged particle on the circular orbit of radius $r$. The corresponding energy $E$ can be found by substituting this value of the angular momentum in the first equation in (\ref{circ}). However, unlike the results for the uncharged particles discussed  above, the expressions obtained are too complicated to be listed here and we plan to return to this subject in future work in order to properly address all the cases for all possible values of the parameters involved \cite{stelea}.

\subsection{Null trajectories}

Finally, let us briefly turn our attention to the null geodesics of the metric (\ref{metric}). For a detailed study of the null geodesics and circular orbits for the Kiselev black hole see \cite{Fernando:2012ue}, \cite{Uniyal:2014paa} and \cite{Malakolkalami:2015cza}.

For the line element (\ref{metric}), with the metric (\ref{fr}) and the conserved quantities (\ref{ELn}), in the equatorial plane (for $\theta=\frac{\pi}{2}$) one obtains the following relation 
\begin{equation}
\dot{r}^2 = E^2 - f(r)\frac{L^2}{r^2}
\end{equation}
pointing out the effective potential
\begin{equation}
V_{eff}(r) =  f(r) \frac{L^2}{r^2} = \left( 1 - \frac{2M}{r} - kr^p \right)  \frac{L^2}{r^2}.
\label{potnull}
\end{equation}

This potential was previously studied in \cite{Malakolkalami:2015cza} for the more general case of the Schwarzschild-de Sitter with quintessence solution. The Kiselev parameters in that work are $c=k>0$ while $\omega=-\frac{p+1}{3}$ in terms of our reparameterization of our metric.

In order to find the null circular orbits, we impose the conditions (\ref{circular}) which lead to the radius values $r_0$ that are solutions of the equation:
\beqs
1-\frac{2M}{r_0}-\frac{2E^2r_0^2}{L^2}&=&\frac{1}{p-2}\left(\frac{6M}{r_0}-2\right).
\eeqs

For example, if $p=1$ then one obtains \cite{Malakolkalami:2015cza}:
\begin{equation}
r_0 = \frac{1 \mp \sqrt{1-6kM}}{k}
\label{r0pm}
\end{equation}
and the following relations between the energy and the angular momentum:
\begin{equation}
\frac{E^2}{L^2} = \frac{1-9kM \pm (1-6kM)^{3/2}}{54 M^2} =
\frac{r_0 -2M -k r^3_0}{r_0^3}.
\label{ELn2}
\end{equation}
For $6kM \ll 1$, one of the circular radius is close to the last Schwarzschild circular orbit $r_0 = 3M$, while the other one, $r \approx (2-3kM)/k$, is not physical since it leads to a negative value for the right hand side of the relation (\ref{ELn2}). Similar results have been previously derived in literature \cite{Fernando:2012ue} - \cite{Malakolkalami:2015cza}. These results depend solely on the characteristics of the original Kiselev geometry, therefore we do not expect any changes in terms of the nonlinear electromagnetic fields in our reinterpreted Kiselev solution.

However, for future work, it would be interesting to consider the behavior of massless fields with electric charges in the background of the nonlinear electromagnetic fields in our solution.

\subsection{Photon orbits for the black hole in the power-Maxwell electrodynamics}

Even if we considered above the null geodesics, one should note that photons behave differently in the context of a nonlinear electrodynamics as compared to the standard linear theory of Maxwell electrodynamics \cite{Novello:1999pg}, \cite{Breton:2007bza}, \cite{Habibina:2020msd}. 

More precisely, photons will follow the null geodesics of an effective geometry given by \cite{Novello:1999pg}:
\beqs
g_{eff}^{\mu\nu}&=&L_Fg^{\mu\nu}-4L_{FF}F^{\mu}_{~\alpha}F^{\alpha\nu},
\label{effmetric}
\eeqs
where $L_{FF}$ is now the second derivative of the power-law Lagrangian with respect to the electromagnetic invariant $F=F_{\mu\nu}F^{\mu\nu}$. In our case $L=-\alpha F^q$ such that $L_F=-\alpha qF^{q-1}$ and $L_{FF}=-\alpha q(q-1)F^{q-2}$. 

Since we are considering the null orbits of the effective geometry, it will suffice to take into account a conformally rescaled effective geometry of the form bellow that will lead to the same null geodesics:
\beqs
g_{\mu\nu}^{resc}&=&g_{\mu\nu}-\frac{4(q-1)}{F}F_{\mu\alpha}F^{\alpha}_{~\nu}.
\eeqs

It is now easy to check that in this case the null geodesics of the rescaled effective geometry satisfy the following equation in the equatorial plane $\theta=\frac{\pi}{2}$:
\beqs
-f(r)\dot{t}^2+\frac{\dot{r}^2}{f(r)}-\frac{p}{2}r^2\dot{\varphi}^2=0,
\label{nullele}
\eeqs
if one uses the electric ansatz for the metric. Formally, taking now into account the conserved quantities (\ref{ELn}) one can readily identify the modified effective potential\footnote{Note that the conserved angular momentum becomes $L=\frac{p}{2}r^2\dot{\varphi}$. The constant factor $p/2$ can be absorbed into the constant $L$, however, the effective potential retains the same form given in (\ref{Veffnullel}).}
\beqs
V_{eff}(r)&=&-\frac{1}{p}\left(1-\frac{2M}{r}-kr^p\right)\frac{L^2}{r^2}.
\label{Veffnullel}
\eeqs
Note that for $0\leq p\leq 2$ the effective potential for the black hole with electric charge has the reversed sign with respect to the corresponding potential in (\ref{potnull}).

If one uses the solution with the magnetic ansatz, the null geodesics of the effective geometry (\ref{effmetric}) satisfy in the equatorial plane $\theta=\frac{\pi}{2}$ a slightly modified equation (as opposed to the electric case):
 \beqs
-f(r)\dot{t}^2+\frac{\dot{r}^2}{f(r)}+(2q-1)r^2\dot{\varphi}^2=0,
\label{nullmag}
\eeqs
which, however, reduces to (\ref{nullele}) once one takes into account (\ref{qmagnetic}). Therefore, in the magnetic case one has the same effective potential (\ref{Veffnullel}).

The effects of this effective potential are strange! There should be a stable circular orbit for photons, while photons with low energies fall directly into the black hole, those with high energies are scattered back to infinity and do not enter the black hole. However, the interpretation of (\ref{Veffnullel}) as an effective potential is not the correct one once one notices the signature of the effective geometry on which photons propagate as in (\ref{nullmag}). Basically, the nonlinear photons will see the radial coordinate $r$ in the static patch as a timelike coordinate. Their trajectories could be integrated directly, however, we will leave this subject for further work \cite{stelea}.

\section{Thermodynamic properties of the Kiselev black hole in power-Maxwell electrodynamics}
\label{section5}

In this section we will initiate the study of the thermodynamics of the Kiselev black holes in the power-Maxwell nonlinear electrodynamics using the electric ansatz from section \ref{section3}. The magnetic case is completely similar. One should contrast the simplicity of the results obtained here with the approach used for instance in \cite{Azreg-Ainou:2014lua}, \cite{Azreg-Ainou:2014twa}, where the effects of the quintessence have been treated by introducing effective pressures and effective volumes, unlike the analysis performed bellow.

As we have seen in section \ref{horizonsg}, this geometry can have at most two horizons, one black hole horizon, located at $r_b$ and one outer, cosmological-type horizon located at $r_c>r_b$. Depending on the values of the parameters, if one increases the mass $M$ then $r_b$ increases while $r_c$ decreases, \textit{i. e.} the two horizons get closer and closer and they will coincide in this coordinate system if the mass reaches the maximum value (\ref{mmax}). If one increases the mass parameter beyond this maximum value the geometry will have no horizon, while it will generically have a naked curvature singularity in origin.

To compute the black hole temperature we will make use of the definition of the surface gravity:
\beqs
k_b^2&=&-\frac{1}{2}\nabla_{\mu}\xi_{\nu}\nabla^{\mu}\xi^{\nu},
\eeqs
where $\xi^{\mu}$ is a Killing vector field which is null on the black hole horizon. Since our metric is static one picks $\xi^{\mu}=\frac{\partial}{\partial t}$ such that the surface gravity becomes $k_b=\frac{f'(r_b)}{2}$ and the black hole temperature becomes:
\beqs
T_b&=&\frac{k_b}{2\pi}=\frac{f'(r_b)}{4\pi}=\frac{p-2^{\frac{2-p}{4}}Q_e^{\frac{2-p}{2}}r_b^p}{4\pi pr_b}
\label{htemp},
\eeqs
where $r_b$ is the radius of the event horizon. For very small values of the black hole horizon radius $r_b$ one can see from (\ref{htemp}) that $T_b\ra\frac{1}{4\pi r_b}$, which is the temperature of a Schwarzschild black hole with horizon radius $r_b=2M$. Note that $r_b$ can only increase up to the maximum value $r_0$ in (\ref{r0}) that corresponds to the extremal case when the black hole and the cosmological horizon coincide. In terms of the electric charge $Q_e$ and the parameter $p$, this maximum value of the black hole horizon can be expressed as:
\beqs
r_0&=&Q_e^{\frac{2-p}{2p}}\left(2^{\frac{p-2}{4}}p\right)^{\frac{1}{p}}.
\label{r0a}
\eeqs
This is precisely the value for which the black hole temperature (\ref{htemp}) reaches zero, as expected.

One can associate as well a temperature to the cosmological horizon, $r_c$:
\beqs
T_c&=&-\frac{k_c}{2\pi}=-\frac{f'(r_c)}{4\pi}.
\eeqs
The minus sign appears here in order to account for the fact that $k_c<0$ on the cosmological horizon.

Note that the entropy of these black holes should satisfy the area-law, which means that the black hole entropy is $S_b=\frac{A_b}{4}$, where $A_b$ is the area of the black hole horizon.  One can also associate an entropy with the cosmological horizon $S_c=\frac{A_c}{4}$, where $A_c$ is the area of the cosmological horizon. The situation here is reminiscent of black holes in de Sitter spacetime. Once again one has a black hole horizon surrounded by an outer, cosmological horizon and the two horizons will coincide in the extremal case, described by the so-called Nariai-de Sitter black hole.

 Similarly to the black hole in the de Sitter case, an important difficulty is associated with the definition of the quasilocal mass for this class of spacetimes. The problem arises here because of the absence of a globally-defined timelike Killing vector at spacial infinity. However, there does exist a Killing vector that is timelike inside the static patch of the Kiselev geometry, while it becomes spacelike outside the cosmological horizon and this vector could still be used to define a notion of quasilocal mass. Another problem is related to the fact that for general values of $p$ the spacetime is not asymptotically de Sitter, nor asymptotically flat and there are no counterterms known to render the action and the conserved physical quantities finite.
 
However, to compute the quasilocal mass one can still make use of the background subtraction method of Brown and York \cite{Brown:1992br}, \cite{Brown:1994gs}. Unlike the counterterm method in de Sitter case \cite{Balasubramanian:2001nb}, this procedure will produce results that depend on the choice of the reference background. Let us begin by writing the metric induced on equal time surfaces (they are the $r=const.$ surfaces outside the cosmological horizon) in the form:
 \beqs
 h_{ab}dx^adx^b&=&-f(r)dt^2+r^2(d\theta^2+\sin^2\theta d\varphi^2)\equiv -f(r)dt^2+\sigma_{ij}d\phi^id\phi^j.
 \label{bound}
 \eeqs
 If $\xi^{\mu}=\frac{\partial}{\partial t}$ is the Killing vector generating an isometry on the boundary and if $n_{\mu}=[\sqrt{-f(r)},0,0,0]$ is the unit normal on a surface of fixed $t$ then, following \cite{Balasubramanian:2001nb} we define the conserved charge associated to the Killing vector $\xi^{\mu}$ using the formula:
 \beqs
 {\cal M}&=&\frac{1}{8\pi}\int_{S^2}d^2\phi\sqrt{\sigma}T_{ab}\xi^an^b,
\label{quasim}
 \eeqs
where we defined \cite{Brown:1992br}, \cite{Brown:1994gs}:
\beqs
T_{ab}&=&(K_{ab}-Kh_{ab})-(K_{ab}^0-K^0h_{ab}^0).
\eeqs
Here $K_{ab}$ is the extrinsic curvature of the metric (\ref{bound}) induced on the boundary, $K$ is its trace, while $K_{ab}^0$, $K^0$ and $h_{ab}^0$ are the corresponding quantities computed for the reference background. This mass formula is used for a surface of fixed time $r$ outside the cosmological horizon, while in the limit in which this boundary is pushed to future infinity $r\rightarrow\infty$ one obtains a finite result for the conserved quasilocal mass.

In our case we shall pick the reference geometry as the one corresponding to $M=0$ in (\ref{fr}). A simple computation leads to a remarkable simple formula for the conserved mass:
\beqs
{\cal M}=\lim\limits_{r\ra\infty}r\sqrt{-f(r)}\big[\sqrt{-f^0(r)}-\sqrt{f(r)}\big],
\eeqs
which can be compared to a similar expression derived in \cite{Chan:1995fr}. Using now the function (\ref{fr}) while $f^0(r)=1-kr^p$, one obtains the quasilocal mass ${\cal M}=M$.

To express the mass ${\cal M}=M$ in terms of the extensive parameters $S_b$ and $Q_e$ we shall use the relation $f(r_b)=0$. One obtains:
\beqs
{\cal M}&=&\frac{r_b}{2p(p+1)}\left(p(p+1)-2^{\frac{2-p}{4}}Q_e^{\frac{2-p}{2}}r_b^p\right).
\eeqs

One can now check that $T_b=\left(\frac{\partial {\cal M}}{\partial S_b}\right)_{Q_e}$ is indeed the Hawking temperature (\ref{htemp}), where $S_b=\pi r_b^2$ is the black hole entropy.

If one defines the electric potential of the black hole:
\beqs
\Phi^e_b&=&\left(\frac{\partial {\cal M}}{\partial Q_e}\right)_{S_b}=\frac{p-2}{2p}\frac{Q_e^{\frac{2-p}{2}}r_b^{p+1}}{2^{\frac{2+p}{4}}(p+1)},
\label{potel}
\eeqs
then it is easy to verify that the first law of thermodynamics is satisfied:
\beqs
d{\cal M}&=&T_bdS_b+\Phi^e_bdQ_e,
\label{first1}
\eeqs
as well as a Smarr like relation of the form:
\beqs
{\cal M}&=&2T_bS_b+\frac{2p}{p-2}\Phi_eQ_e.
\label{smarr1}
\eeqs
This is precisely the same Smarr relation found in \cite{Gonzalez:2009nn}, as expected. Note that there are similar relations corresponding to the cosmological horizon $r_c$:
\beqs
d{\cal M}&=&-T_cdS_c+\Phi^e_cdQ_e,
\label{first2}
\eeqs
where 
\beqs
\Phi^e_c&=&\left(\frac{\partial {\cal M}}{\partial Q_e}\right)_{S_c}=\frac{p-2}{2p}\frac{Q_e^{\frac{2-p}{2}}r_c^{p+1}}{2^{\frac{2+p}{4}}(p+1)},
\label{potelc}
\eeqs
while
\beqs
{\cal M}&=&-2T_cS_c+\frac{2p}{p-2}\Phi^e_cQ_e.
\label{smarr2}
\eeqs

We are now ready to investigate the thermal stability using the canonical ensemble method\footnote{There are also stability conditions that can be checked at the level of the nonlinear electrodynamics Lagrangian \cite{Moreno:2002gg}.}. We shall compute the black hole heat capacity while keeping the black hole charge $Q_e$ as constant. The heat capacity becomes:
\beqs
C_{Q_e}&=&T_b\left(\frac{\partial {S_b}}{\partial T_b}\right)_{Q_e}=\frac{T_b}{M_{SS}},
\label{heatq}
\eeqs
where we defined $M_{SS}=\left(\frac{\partial^2 {\cal M}}{\partial S_b^2}\right)_{Q_e}$. Positivity of $C_{Q_e}$ or $M_{SS}$ would be sufficient to ensure the local stability of our black holes in the power-Maxwell electrodynamics. In our case $T_b$ is positive and it reaches the zero value for $r_b=r_0$ from (\ref{r0a}). Therefore, one should turn our attention to the quantity:
\beqs
M_{SS}&=&-\frac{p+(p-1)2^{\frac{2-p}{4}}Q_e^{\frac{2-p}{2}}r_b^p}{8\pi^2 pr_b^3}.
\eeqs
Since $p\in[0,2]$ there might be values for $r_b$ for which $M_{SS}=0$ if $0\leq p\leq 1$, which generically will lead to divergences in the heat capacity $C_{Q_e}$, signaling type $2$ phase transitions. However, if one recalls the expression of $r_0$ from (\ref{r0a}) then one can express the quantity $M_{SS}$ as:
\beqs
M_{SS}&=&-\frac{1-(1-p)\left(\frac{r_b}{r_0}\right)^p}{8\pi^2 r_b^3}.
\label{mss}
\eeqs
It is now clear that if $0\leq p\leq 1$ then $M_{SS}$ can have a real root $r_b=r_0(1-p)^{-\frac{1}{p}}$, however this value is always greater than the maximum value $r_0$ that can be reached by the black hole horizon. In conclusion, in the interval $0\leq r_b\leq r_0$ the heat capacity is always negative and the black hole system in the power-Maxwell electrodynamics is unstable for $p\leq 2$. The point $r_b=r_0$ is a bound point of the heat capacity, since for this value one has $C_{Q_e}=T_b=0$. However, at this location the heat capacity does not change sign and there are no type $1$ phase transitions since there are no physical values of the black hole horizon radius $r_b$ beyond the limit $r_0$ given in (\ref{r0a}) for which the heat capacity $C_{Q_e}$ could reach positive values.

\begin{figure}
  \centering
  \includegraphics[width=0.45\textwidth]{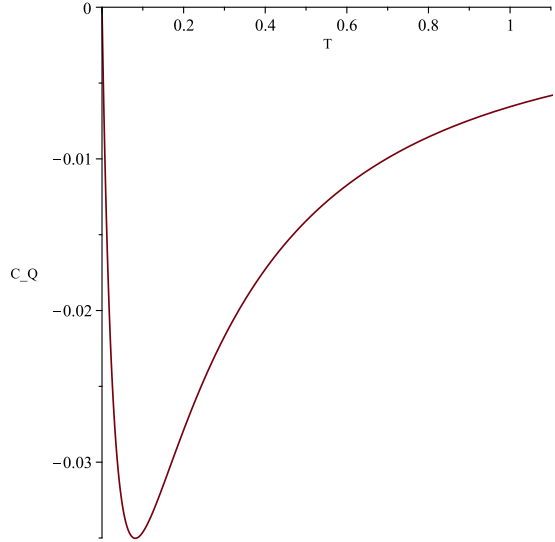}
  \caption{The heat capacity, $C_{Q_e}$ versus the temperature $T_b$ for $Q_e=0.5$ and $p=\frac{2}{5}$. One can notice the presence of a Schottky peak.} 
  \label{figure5}
\end{figure}

The black hole temperature $T_b$ varies from $0$ to $\infty$. The value $T_b\ra\infty$ is attained in the limit $r_b\ra 0$. From (\ref{heatq}) and (\ref{mss}) it should be clear that in this limit $C_{Q_e}\ra 0$. As shown in Figure \ref{figure5} this signals the presence of a Schottky peak \cite{Dinsmore:2019elr}, \cite{Johnson:2019ayc}, \cite{Johnson:2019vqf} in the dependence of temperature of the heat capacity (\ref{heatq}). 

With hindsight, the presence of a Schottky peak was to be expected in our case. They usually appear in multi-horizon spacetimes (although they have been noticed in the anti-de Sitter case as well \cite{Johnson:2019vqf}), such as black holes in de Sitter geometry, for which there is a cap in the energy of the system. In this case the dependence of the heat capacity as a function of temperature could give us important clues regarding the underlying degrees of freedom for such systems. One should note that, similar to the Schwarzschild - de Sitter case, the existence of the Schottky peaks is directly related to the existence of a cosmological horizon \cite{Dinsmore:2019elr}.

Furthermore, the existence of the Schottky peak hints to the intriguing possibility of using the black hole in the power-Maxwell electrodynamics to function as a continuous heat engine \cite{Johnson:2019ayc}. More precisely, combining (\ref{first1}) with (\ref{first2}) one obtains:
\beqs
2d{\cal M}&=&T_bdS_b-T_cdS_c+(\Phi^e_b+\Phi^e_c)dQ_e.
\eeqs
The continuous heat engine mode of operation will leave the black hole energy fixed $d{\cal M}=0$ such that:
\beqs
T_bdS_b&=&T_cdS_c-(\Phi^e_b+\Phi^e_c)dQ_e.
\eeqs
Consider now the case with $dQ_e>0$, with an increase of the black hole entropy $dS_b>0$ such that there is a positive heat inflow $Q_H=T_bdS_b>0$, with $Q_C=T_cdS_c$ the heat flow away from the black hole. Then $W=-(\Phi^e_b+\Phi^e_c)dQ_e>0$ is the positive work done in this cycle.

\section{Conclusions}
\label{section6}

In recent years, Kiselev's solution has received increased interest in connection to the properties of the anisotropic fluid sourcing this geometry, properties that mimic a dark energy source. In the present work we found a physical source for the Kiselev geometry within the context of nonlinear electrodynamics theories. More specifically, in section \ref{section3} we showed that the Kiselev geometry becomes an exact solution of the Einstein equations coupled to power-Maxwell electrodynamics, using either an electric or a magnetic ansatz. We also checked the energy conditions showing that the weak energy condition (WEC) and the dominant energy condition (DEC) are both satisfied, while the strong energy condition (SEC) is violated in both cases.

In section \ref{section4} we studied the trajectories of charged and uncharged particles that follow the timelike and the null geodesics of this geometry. However, in the nonlinear electrodynamics theories it turns out that photons will not follow the null geodesics of the background Kiselev geometry, instead they move along the null geodesics of an effective geometry, defined in (\ref{effmetric}). As it appears, the nonlinear photons will see the radial coordinate $r$ in the static patch as a timelike coordinate, while the time coordinate $t$ becomes spacelike (just as these coordinates change their roles when crossing a horizon). Their trajectories could be integrated directly, however, we will leave this subject for further work \cite{stelea}.

In section \ref{section5} we investigated the thermodynamics of the Kiselev solution in the power-Maxwell geometry for the electric case. We computed the mass using the Brown-York subtraction method and showed that the first law of thermodynamics and a modified Smarr relation are both satisfied. The black hole in this case has negative heat capacity and is therefore unstable, while it exhibits a Schottky peak in the dependence of the heat capacity as a function of temperature. Similarly to the de Sitter case, the presence of the Schottky peak hints to the possibility of using this black hole as a continuous heat engine.

As avenues for further work, it might be interesting to investigate the effects of nonlinear power-Maxwell fields in constructing interior fluid solutions, which describe compact objects in General Relativity. Such solutions can be generated easily for the usual Maxwell electrodynamics \cite{Stelea:2018cgm}, \cite{Stelea:2018elx} and it might be fruitful to further investigate this matter. 

Another interesting issue is the the study of the behavior of charged scalar and spinorial fields in the background of the reinterpreted Kiselev geometry in the nonlinear electrodynamics. Following similar analysis performed in \cite{Dariescu:2019psb} - \cite{Dariescu:2018dyy}, it is quite possible that for particular values of the parameter $p$ the solutions can be expressed analytically by means of the Heun functions \cite{Heun}. Work on these issues is in progress and it will be reported elsewhere.

\section*{Acknowledgements}

The authors would like to thank the anonymous referees whose remarks and suggestions helped improve this manuscript.

\end{document}